\documentclass[12pt]{iopart}

\usepackage{amssymb}
\usepackage{iopams}
\usepackage{graphicx}

\def\anta{{\sc Antares}}
\def\vo{{\sc Virgo}}
\def\lo{{\sc LIGO}}

\begin{document}




\title{The \anta~Neutrino Telescope and Multi-Messenger Astronomy}


\author{Thierry PRADIER$^{*}$, on behalf of the \anta~Collaboration\footnote{\texttt{http://antares.in2p3.fr}}}
\address{$^{*}$ University of Strasbourg (France) \& {\sl Institut Pluridisciplinaire Hubert Curien}}

\begin{abstract}
\anta~is currently the largest neutrino telescope operating in the Northern
Hemisphere, aiming at the detection of high-energy neutrinos from astrophysical sources. Such
observations would provide important clues about the processes at work in those sources, and possibly help solve the puzzle of
ultra-high energy cosmic rays. In this context, \anta~is developing several programs to improve its capabilities of
revealing possible spatial and/or temporal correlations of neutrinos with other cosmic messengers: photons, cosmic rays and gravitational waves.
The neutrino telescope and its most recent results are presented, together with these multi-messenger programs.
\end{abstract}

%

\section{Introduction}
\label{sec:intro}
Astroparticle physics has entered an exciting period with
the recent development of experimental techniques that have
opened new windows of observation of the cosmic radiation
in all its components: photons, cosmic rays, but also gravitational waves and high energy
neutrinos, that could be detected both by IceCube \cite{icecube} and \anta.


The advantage of using neutrinos as new messengers lies firstly on their weak interaction cross-section ; 
unlike protons or $\gamma$, they provide a cosmological-range 
unaltered information from the very heart of their sources. 
Secondly, charged particles are deflected by magnetic fields. Neutrinos on the other hand point directly to their sources and exact production site.

The neutrinos \anta~is aiming at are typically TeV neutrinos from AGNs or galactic sources (microquasars), 30 orders of magnitude lower in flux than solar neutrinos. 
The detection of those specific neutrinos requires under water/ice instruments, 
or alternatively acoustic/radio techniques in the PeV-EeV range and air showers arrays above 1 EeV.
In spite of efforts in those various energy ranges, since the detection of the MeV neutrino burst from SN 1987A by {\sc Kamiokande/Baksan/imb/Mont-Blanc} \cite{sn1987} no astrophysical source for neutrinos above a few GeV has ever been identified.

Sources for TeV $\nu$ are typically compact objects (neutron stars/black holes), from which often emerge relativistic plasma jets with a still unclear 
composition - purely leptonic or with some hadronic component.
Most of these sources have already been extensively studied from radio wavelengths up to $\gamma$-rays. 
%
These photons can be produced by $e^-$ {\it via} inverse compton effect (on ambient photon field)/synchrotron radiation, or by protons/nuclei {\it via} 
photoproduction of $\pi^{0}/\pi^{\pm}$ :

\begin{equation}
\begin{array}{l}
 p / A  +  p / \gamma \longrightarrow \pi^{0}~\pi^{\pm} \\
\textrm{where~}\pi^{0} \rightarrow \gamma \gamma~\textrm{and}~\pi^{\pm} \rightarrow \nu_{\mu}~\mu, \textrm{~with~} \mu \rightarrow \nu_{\mu} \nu_e e \\
\end{array}
\label{eq:prodmu}
\end{equation}

\begin{figure}[h!]
\centerline{\includegraphics[width=\linewidth]{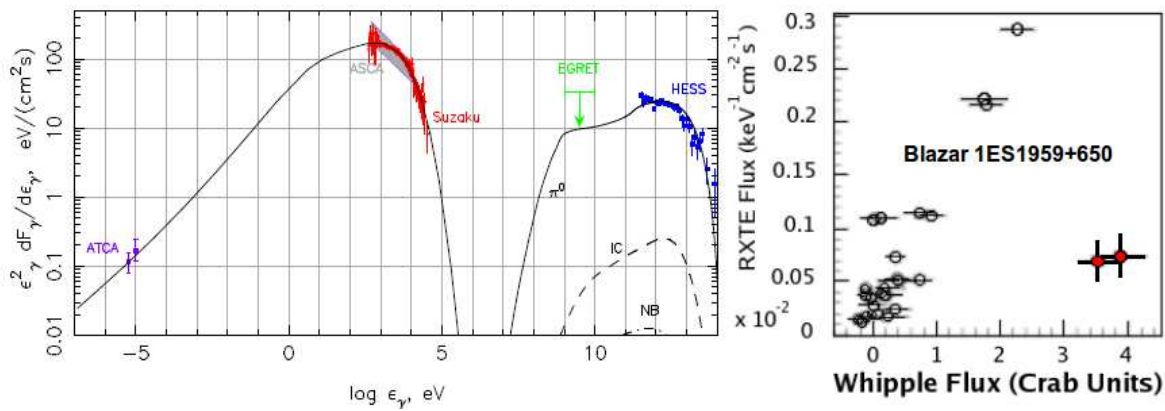}}
\caption{Left: Multiwavelength observations of the SNR RXJ 1713.7-39; the solid curve at energies above $10^7$~eV
  corresponds to $\pi^0$-decay $\gamma$-ray emission, whereas the dashed and dash-dotted
  curves indicate the inverse Compton (IC) and Nonthermal Bremsstrahlung (NB)
  emissions, respectively. Right: Whipple vs RXTE flux, for the Blazar 1ES1959+650, which shows the existence of orphan $\gamma$ flares (crosses).
\label{fig1}}
\end{figure}

In the former scenario, no neutrinos are produced, whereas in the latter, the neutrino flux is directly related to the gamma flux: a TeV neutrino detection 
from gamma sources would then yield a unique way to probe the inner processes of the most powerful events in the universe. Several hints exist which indicates 
that hadrons could be accelerated up to very high energies. Firstly, the combined radio, X-rays and $\gamma$-rays observations of the shell-type supernova 
remnant RX J1713.7-3946 \cite{rx} favour the production of photons {\it via} $\pi^0$ decay (figure \ref{fig1}, left). Secondly, the correlations 
between X and $\gamma$ for the Blazar 1ES1959+650 \cite{1es} prove the existence of $\gamma$ flares not visible in X (figure \ref{fig1}, right), 
which is difficult to account for in purely leptonic models. 

The connection between high energy neutrino astronomy and both gamma-ray astronomy, charged cosmic rays and gravitational waves thus emphasizes 
that not only a multi-wavelength but also a multi-messenger approach, combining data from different observatories, is suited for 
the study of the most powerful astrophysical sources in the Universe.

Sections \ref{sec:telescope} and \ref{sec:results} first describe the \anta~neutrino telescope and its latest physics results. Section \ref{sec:grbs} presents the different
strategies imagined to detect neutrinos from gamma-ray bursts, while section \ref{sec:correlations} details the correlations that can be performed
with other observatories, such as {\sc Auger}, {\sc Hess}, or \vo/\lo.

%

\section{The \anta~Neutrino Telescope: principles \& description}
\label{sec:telescope}

\anta~is a three-dimensional grid of photomultiplier tubes, arranged in strings, able to detect the
\v{C}erenkov photons induced by the passage in sea water of relativistic charged particles, produced by the interaction of a cosmic neutrino in the Earth \cite{markov}.
The measurements of the time of the hits (with a time resolution of the order of ns) and the amplitude of the hits (with a resolution of about  $30 \%$), 
together with the position of the hits (by measuring the position of each PMT, to reach a resolution of about 10 cm) are needed to achieve the 
reconstruction of those signals with the desired resolution. Muon tracks initiated by the charged current interaction of a $\nu_{\mu}$ in the Earth are detected {\it via} their directional \v{C}erenkov light 
and can be reconstructed with an angular resolution below 0.3$^{\circ}$ above 10 TeV. The resolution below this energy is dominated by the kinematics of the interaction.
 The energy resolution is quite poor, a factor 2-3 on average, restricted by the granularity/density of the light sensors and the fact that the muon traverses the 
detector. Showers produced by $\nu_e$ on the other hand emit quasi-isotropic light, and can be reconstructed with a better energy resolution (roughly 30 \%) but 
with a poorer angular resolution, typically~3-5~$^{\circ}$.

The main physical backgrounds are twofold. Atmospheric muons produced in the upper atmosphere
by the interaction of cosmic rays can be strongly suppressed because of their downward direction. Upward-going atmospheric neutrinos on the other hand are 
more delicate to identify: they have exactly the same signature as the expected cosmic signal \anta~awaits for.

\begin{figure}[h!]
\centerline{\includegraphics[width=0.5\linewidth]{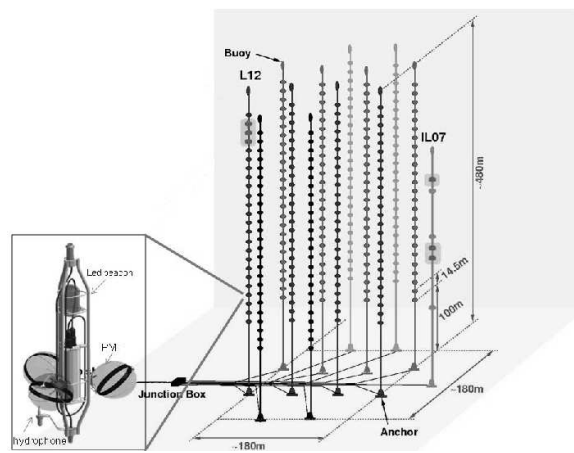}}
\caption{The layout of the completed \anta~detector.
Bottom left: optical storey with its three photomultipliers.
\label{fig2}}
\end{figure}

The \anta~neutrino telescope, deployed at 2500 m below sea surface, 40 km off the coast of Toulon (Southern France) is composed of 12 strings, with 25 
storeys each containing a triplet of 10$^{``}$ photomultipliers oriented at 45 degrees downward to be optimally sensitive to upward going muons \cite{antares}. Since May 2008, 12 lines
 are continuously taking data. A schematic description of the detector, together with the layout of the lines, and of one of the storey, 
can be found in figure \ref{fig2}. 

\section{Selected results from \anta}
\label{sec:results}

The main results obtained recently, related to atmospheric muons and neutrino-induced upward-going muons, are presented in this section.
\subsection{Atmospheric muons}

\begin{figure}[h!]
\centerline{\includegraphics[width=\linewidth]{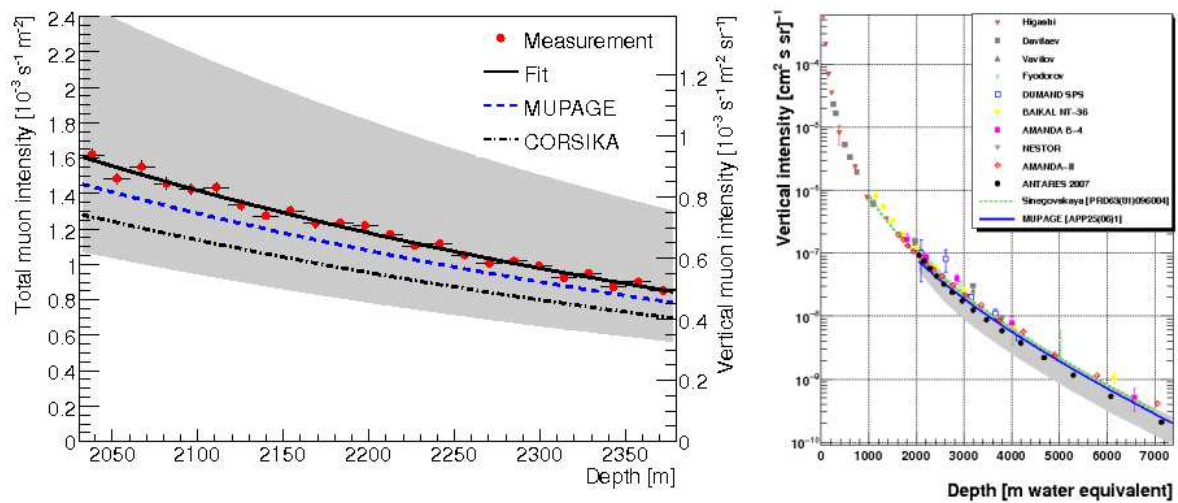}}
\caption{Left: Attenuation of the flux of muons as a function of
depth, as extracted using the method described in \protect\cite{anta_muon}. Right: Vertical depth intensity relation of atmospheric
muons with $E_{\mu} > 20~GeV$ (black points).
\label{fig3}}
\end{figure}

The attenuation of the muon flux as a function
of depth can be observed in figure~\ref{fig3}~(left), as computed using the method described in \cite{anta_muon}.
Alternatively, the reconstructed zenith angle can be converted
to an equivalent slant depth through the sea water, and a depth intensity relation can
be extracted, also shown in figure~\ref{fig3}~(right). The results are in agreement with previous measurements.

\subsection{Neutrinos}

\begin{figure}[h!]
\centerline{\includegraphics[width=\linewidth]{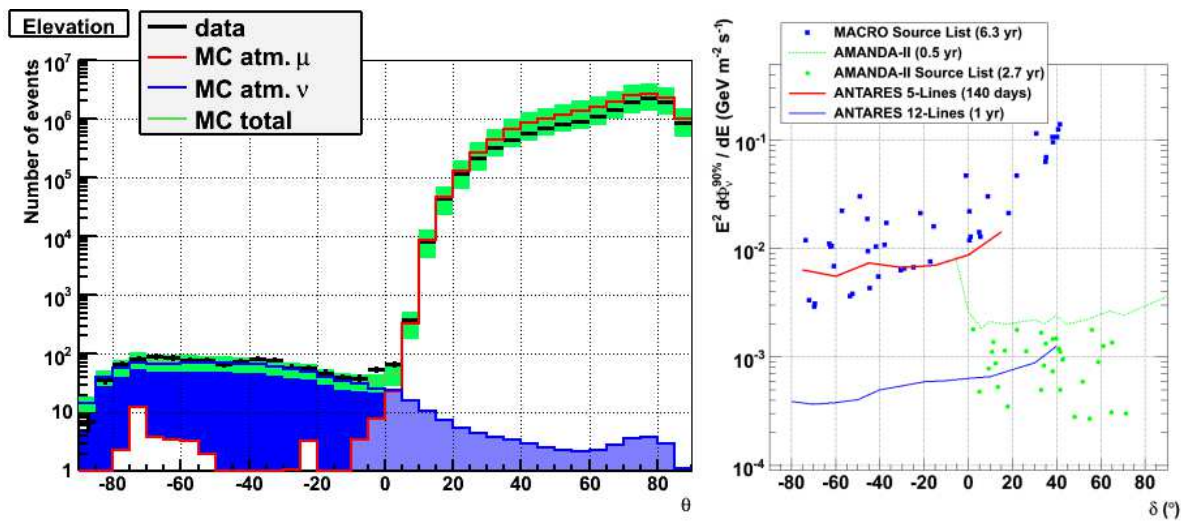}}
\caption{Left: Zenith distribution of reconstructed muons in the
2007+2008 data. Right: expected sensitivity for \anta~in the 5 line configuration. The expected sensitivity of \anta~for one
year with twelve lines is also shown.
\label{fig4}}
\end{figure}

The muons produced by the interaction of neutrinos
can be isolated from the atmospheric muons by requiring that the muon trajectory is
reconstructed as up-going. In figure \ref{fig4} the zenith angular
distribution of muons in the 2007+2008 data (5 lines and 9-12 lines) sample is shown. A total of 1062
up-going neutrino candidates are found, in
good agreement with expectations from the atmospheric
neutrino background.

An all sky search, independent of assumption on the source
location, has been performed on the 5-line data. No significant cluster was found. 
A search amongst a pre-defined list of 24 of the most
promising galactic and extra-galactic neutrino sources
({\it e.g.} supernova remnants, BL Lac objects) was performed. The
corresponding flux expected sensitivities, assuming an $E^{−2}$ flux,
are also plotted in figure \ref{fig4} and compared to published upper
limits from other experiments (see \cite{amanda} for {\sc Amanda}). Also shown is the predicted upper limit for \anta~after one full
year of twelve line operation. 

\section{Searches for neutrinos from GRBs}
\label{sec:grbs}


Among all the possible astrophysical sources, transient sources
offer one of the most promising perspectives for
the detection of cosmic neutrinos thanks to the almost
background free search. For instance, several models predict the production of high-energy
neutrinos by gamma-ray bursts (GRBs) \cite{grbnu}: the detection
of these neutrinos would provide evidence for
hadron acceleration by GRBs. 

Two different methods to detect transient sources can be used.
The \textit{triggered search} is based on the search for neutrino
candidates in conjunction with an accurate timing and
positional information provided by an external source. The \textit{rolling search} is based
on the search for high energy or multiplet of neutrino
events coming from the same position within a given
time window.

\subsection{Triggered search with the GCN}

\begin{figure}[h!]
\centerline{\includegraphics[width=\linewidth]{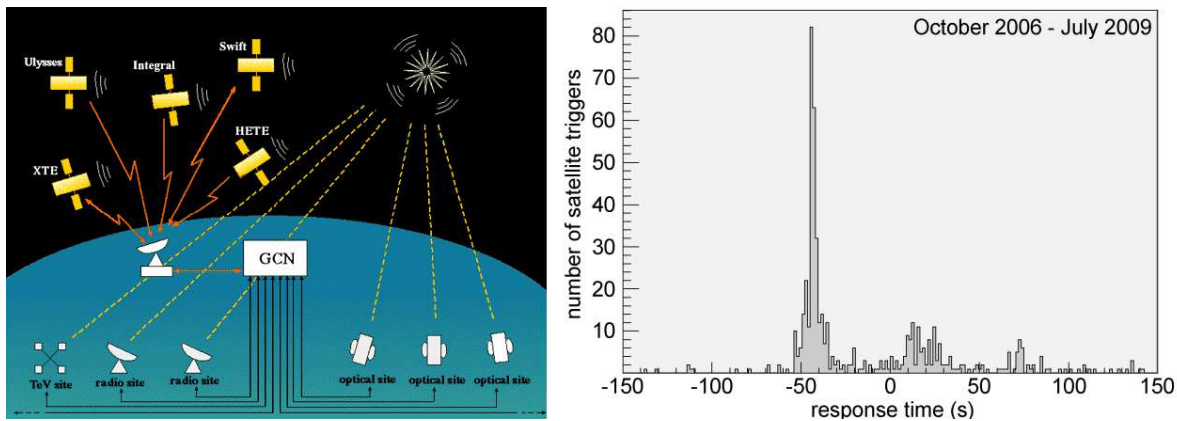}}
\caption{Left: the Gamma-ray bursts Coordinates Network (GCN). Right: \anta~response time to an alert.
\label{fig5}}
\end{figure}

Classically, GRBs or flare of AGNs are detected by gamma-ray satellites (Swift, Fermi)  which deliver in real time an alert
to the Gamma-ray bursts Coordinates Network (GCN, figure \ref{fig5}). The characteristics (direction and time of the detection) 
of this alert are then distributed to the other observatories. The small difference in arrival
time and position expected between photons and neutrinos allows a very efficient detection by reducing
the associated background. This method has been implemented in \anta~mainly for the GRB detection
since the end of 2006. Data triggered by more than 500 alerts have been stored up to now. 

Based on the time of the external alert, in complement to the
standard acquisition strategy, an on-line running program stores the data coming from the whole detector during 2
minutes without any filtering. This allows to lower the energy threshold of the event selection during the offline
analysis with respect to the standard filtered data. Due to a continuous buffering of data (covering 60s)
and thanks to the very fast response time of the GCN network (see figure \ref{fig5}), \anta~is able to record data before the
detection of the GRB by the satellite \cite{icrc_mieke}. The analysis of the data relying on those external alerts is on-going.

Due to the very low background rate, even the detection of a small number of neutrinos correlated with
GRBs could set a discovery. But, due to the relatively small field of view of the gamma-ray satellites (\textit{e.g.}, Swift has a 1.4 sr field of view), 
only a small fraction of the existing bursts are triggered. Moreover, choked GRBs without photons counterpart can not
be detected by this method. This justifies the use of a complementary {\it rolling} search strategy.

\subsection{Rolling search: the {\sc TAToO} project}

\begin{figure}[h!]
\centerline{\includegraphics[width=\linewidth]{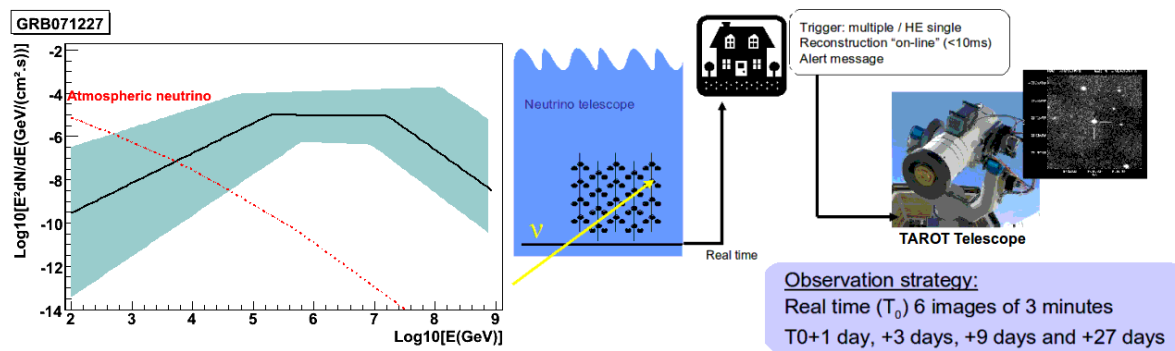}}
\caption{Left: a typical GRB neutrino spectrum. Right: Concept of {\sc TAToO}.
\label{fig6}}
\end{figure}

This second method relies on the detection of a burst of neutrinos in temporal and directional coincidence.
Applied to \anta, the detection of 2 neutrinos within a short time is almost statistically significant: the number of doublets due to atmospheric neutrino 
background events is of the order of 0.004 per year when a temporal window of 900~s and a directional cone of~$2^{\circ}\times2^{\circ}$ are defined. 
It is also possible to search for single cosmic neutrino events by requiring that the reconstructed muon energy is higher than a given energy
threshold (typically above a few tens of TeV), for which the atmospheric neutrino background is negligible (see figure \ref{fig6}). 
When
the neutrino telescope is running, this method is almost 100\% efficient and applies whenever
the neutrinos are emitted with respect to the gamma flash. 
The main drawback is that a
detection is not automatically associated to an astronomical source. It is thus fundamental
to organize a complementary follow-up program to confirm the detection.

In this context, \anta~is organizing a follow-up program in collaboration with TAROT (T\'elescope \`a Action Rapide
pour les Objets Transitoires, Rapid Action Telescope for Transient Objects), called {\sc TAToO} (TAROT-\anta~Target of Opportunity, figure \ref{fig6}). 
The TAROT network is composed of
two 25 cm optical robotic telescopes located at Calern (South of France) and La Silla (Chile). The main
advantages of the TAROT instruments are the large field of view of $1.86^{\circ}\times1.86^{\circ}$ and their very fast
positioning time (less than 10 s).
A GRB afterglow requires a very fast observation strategy in contrast to a core-collapse
supernovae for which the optical signal will appear several days after the neutrino signal. The observational strategy is composed of a real time observation 
followed by few observations during the following month. Depending on the neutrino trigger settings, an alert sent to TAROT is issued at a rate of about one or 
two per month. The total latency is quite impressive: less than 10 ms/event for reconstruction, with a pointing accuracy of the order of $0.6^{\circ}$ above 10 TeV, 
less than 1s for alert sending, with a positioning of the telescopes in less than 10 seconds.
A confirmation by an optical telescope of a neutrino alert will not only give the nature of the
source but also allow to increase the precision of the source direction determination in order to trigger other
observatories. The {\sc TAToO} program is operational since February 2009, and several neutrino alerts have 
already been sent~\cite{icrc_dornic}. A torough analysis of these alerts is underway.
\section{Correlations with other observatories}
\label{sec:correlations}

As already pointed out, high energy cosmic neutrinos could be produced in astrophysical sources that are also potential
emitters of charged cosmic rays, gamma-rays or gravitational waves. Correlations with dedicated instruments could then
bring a harvest of unique information about these sources.

\subsection{{\sc Auger} and ultra-high energy cosmic rays}

The Pierre Auger Observatory reported an anisotropy in
the arrival directions of ultra-high energy cosmic rays (UHECR) \cite{auger}. Correlation with
Active Galactic Nuclei (AGN) from the V\'eron-Cetty catalog
was the most significant
for cosmic rays with energies higher than 57 EeV and
AGNs at distances less than 75 Mpc. The anisotropy was estimated with a confidence
level of 99\%. The suggested correlation with
local AGN sources mostly follows the location of the
supergalactic plane.
%
On
the other side, no significant excess above the atmospheric
neutrino flux was reported by the neutrino telescopes IceCube and \anta.

Instead of searching for such a localized excess, neutrino
arrival direction can be correlated with the arrival directions
of ultra high energy cosmic rays, as described in~\cite{auger_petrovic}.
By obtaining the probability density
function of the number of neutrino events within specific
angular distance from observed UHECRs, the number of
neutrino events in the vicinity of observed ultra-high energy
cosmic rays, necessary to claim a discovery with a
chosen significance, can be calculated.
For example, for 27 UHECR,
a correlation significance of 5$\sigma$ is reached with 2\%-
25\% of neutrino events falling in 1-10$^{\circ}$ bins around the original UHECR direction.
Possible observed correlation of the arrival directions of
those two messengers would provide a strong
indication of hadronic acceleration theory.


\subsection{TeV $\gamma$ sources observed by {\sc Hess}}

The exact composition of the emission of the sources observed by {\sc Hess} and other gamma-ray observatories is still an open issue. 
Some of these sources could have a non-negligible hadronic component, and could then be neutrino emitters. 
In the case of continuous sources (as opposed to transients), a stacking analysis can be performed, which consists in summing up all the events detected in a 
given angular region around the direction of the $\gamma$ source, when below the horizon. Extended sources can be divided in different regions of interest to increase the signal-to-noise ratio. 
Obviously, adding more and more sources to the analysis
also increases the number of background events detected. It is found that an optimum of 25 sources is needed to reach a significance close to 3$\sigma$ after 5 years
of \anta~data taking, assuming that all {\sc Hess} sources are hadronic, after which the sensitivity decreases.

There have also been many discussions about the
possibility to detect muons produced by high energy
gamma-rays in underground, underice or underwater
neutrino telescopes. In contrast to upward-going
muons from neutrinos, downward-going muons
from gamma rays suffer from a high atmospheric muon
background. Therefore the sensitivity of a neutrino telescope
to gamma ray induced muons is quite lower than
atmospheric \v{C}erenkov telescopes. However it monitors
continuously all directions. 
There are at least three processes by which a photon
can produce muons : photoproduction, muon pair production
and charm decay. 
It is then possible to look for a global excess of the muon
flux in a direction correlated with the position of the source
(and within a given time window, if the source is transient). If the statistics is
sufficient, which depends on the source spectral index and flux, accurately measured for a number of galactic sources, 
calibration of both the absolute pointing and the angular resolution of the neutrino telescope could
be performed \cite{icrc_goulven}.

\subsection{Gravitational Waves: the {\sc gwhen} project}

Coincident searches of high-energy neutrinos (HEN) and gravitational
waves (GW) are also of great interest and are detailed in \cite{gwhen_eric}. 
Both GW and HEN are alternative cosmic messengers that carry information 
from the innermost regions of the astrophysical engines. Such messengers could also reveal new, hidden sources that were not observed by conventional 
photon-based astronomy.
The observation of such coincidence would be a landmark event and can also provide observational evidence that GW and HEN originate from a common
 astrophysical source. 
%

\begin{figure}[h!]
\centerline{\includegraphics[width=\linewidth]{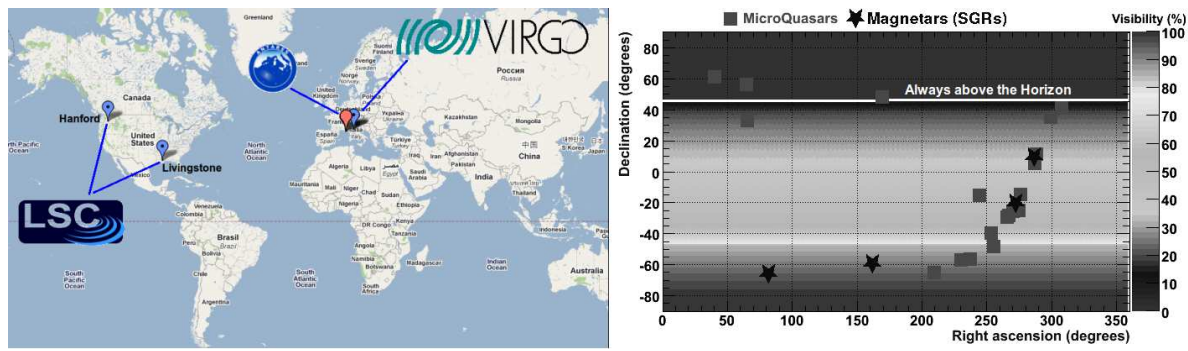}}
\caption{Left: the network of groundbased GW interferometers. The location of \anta~is also shown. Right: averaged visibility skymap for \anta/\vo+\lo, where the location of known microquasars/soft-gamma repeaters, 
potential emitters of GW bursts and HEN, is also shown. 
This daily (normalized) averaged visibility takes into account the interferometers' beam patterns, and the fact that a neutrino source can be detected only when below the horizon.
\label{fig7}}
\end{figure}

The network of GW detectors formed by the \lo~and \vo~interferometers (figure \ref{fig7}) can determine the direction/time of GW bursts in connection with neutrino
events observed in \anta. The \vo/\lo~network started a data-taking phase mid-2009. It monitors roughly 30\% of
the sky in common with \anta:  figure \ref{fig7} shows the daily averaged visibility of some sources by the network. Joint searches are ongoing within a dedicated {\sc gwhen}
working group: investigations \cite{gwhen_titi} led for \anta~indicate that a joint optimized \vo+\lo/\anta~analysis is needed, in order to increase the coincident detection efficiency,
while keeping the coincident false-alarm rate as low as possible, to enhance the significance of such a detection.

\section{Conclusions}
\label{sec:conclu}

\anta~is now taking data since 2008 and has demonstrated the possibility to operate and get competitive physics results from a neutrino telescope under the sea
in the Northern Hemisphere, in spite of its reduced size with respect to IceCube. It should also be noted that, because of its location, \anta~can 
observe the Galactic Centre and most of {\sc Hess} sources on the galactic plane.

Through the online and offline multi-messenger programs described hereabove, the \anta~detector not only enhances
its own capabilities as a neutrino telescope, but also contributes to the global effort of understanding the most violent phenomena
in our Universe. In addition to offline searches for spatio-temporal correlations with other cosmic messengers (photons,
cosmic rays and gravitational waves), \anta~has the capability to handle external alerts in real time and to trigger
follow-up observations with the small latency time required for the study of transient sources. The possibility to store a few
minutes of raw data in coincidence with a GCN alert also brings new opportunities for offline analysis. This could be extended
in the future to handle alerts involving other messengers, such as gravitational waves, to complete the {\sc gwhen} project. 

Finally, the extension of the follow-up programs to other instruments in different ranges of wavelengths (X, radio) would undoubtely contribute to the development 
of the astrophysical potential of \anta.

\vskip 1cm
\noindent
\textbf{Aknowledgements:} \textit{Great thanks to the Organizing Committee, especially Fulvio Ricci, for trusting 
me for this talk.}


\vskip 0.5cm
\begin{thebibliography}{16}
%
\bibitem{icecube} F. Halzen, for {\sc IceCube}, Proceedings of {\sc GWDAW14} (2010)
\bibitem{sn1987} K. S. Hirata {\it et al.}, {\it Phys. Rev. Lett.}~{\bf 58}~(1987)~1490; R. M. Bionta {\it et al}, {\it Phys. Rev. Lett.}~{\bf 58}~(1987)~1494 ; E. N. Alexeyev {\it et al}, {\em PZETF}~{\bf 45}~(1987)~461; {\sc Mont-Blanc} (controversial)~:  V. L. Dadykin {\it et al.}, {\em PZETF}~{\bf 45}~(1987)~464
\bibitem{rx} E. G. Berezhko \& H. J. V\"{o}lk, \texttt{arXiv:0707.4647v1}, 30$^{\textrm{th}}$ {\sc icrc} 2007, Mexico
\bibitem{1es} J. Holder ({\sc Veritas} Collaboration), \texttt{arXiv:astro-ph/0305577v1}, 28$^{\textrm{th}}$ {\sc icrc} 2003, Japan
\bibitem{markov} M. A. Markov, {\em International Conference on High-Energy Physics} 578 (1960)
\bibitem{antares} \anta~Collaboration, {\it Nuclear Instruments and Methods in Physics Research} {\bf A} 555 (2005)~132
\bibitem{anta_muon} \anta~Collaboration, {\it Astropart. Phys.} {\bf 33} (2010) 86-90
\bibitem{amanda} {\sc Amanda} Collaboration, {Phys. Rev. D} {\bf 76} (2007) 042008
\bibitem{grbnu} {\it e.g.} S. Razzaque {\it et al.}, {\it Phys. Rev. D} \textbf{69} (2004) 023001 
\bibitem{icrc_mieke} M. Bouwhuis, for \anta, Proceedings of 31$^{\textrm{st}}$ ICRC conference, Lodz (2009) \texttt{arXiv:astro-ph/0908.0818}
\bibitem{icrc_dornic} D. Dornic, for \anta, Proceedings of 31$^{\textrm{st}}$ ICRC conference, Lodz (2009) \texttt{arXiv:astro-ph/0908.0804}
\bibitem{auger} {\sc Pierre Auger Collaboration}, {\it Science} {\bf 318} (2007) 938
\bibitem{auger_petrovic} J. Petrovic, \texttt{arXiv:astro-ph/0908.1235}
\bibitem{icrc_goulven} G. Guillard for \anta, Proceedings of 31$^{\textrm{st}}$ ICRC conference, Lodz (2009) \texttt{arXiv:astro-ph/0908.0855}
\bibitem{gwhen_eric} E. Chassande-Mottin, for \anta/\lo/\vo, Proceedings of {\sc GWDAW14} (2010)
\bibitem{gwhen_titi} Th. Pradier, {\it Nuclear Instruments and Methods in Physics Research} {\bf A} 602-1 (2009) 268-274

\end{thebibliography}
\end{document}